\documentclass[]{spie}  %>>> use for US letter paper
\usepackage{graphicx}
\usepackage{rotating}
\usepackage{amssymb}
\usepackage{mathptmx}
\usepackage{epstopdf}
\usepackage[]{natbib}
\usepackage{lineno}
\usepackage{url}

\bibpunct{[}{]}{,}{n}{}{,}

\begin{document}

\title{Commercially fabricated antenna-coupled Transition Edge Sensor bolometer detectors for next generation Cosmic Microwave Background polarimetry experiment}

\author{Aritoki Suzuki$^{*a}$ ,  
Nicholas Cothard$^{d}$ , 
Adrian T. Lee$^{a,b}$ , 
Michael D. Niemack$^{d}$ , 
Christopher Raum$^{b,c}$ , 
Mario Renzullo$^{e}$ , 
Trevor Sasse$^{b}$ , 
Jason Stevens$^{d}$ , 
Patrick Truitt$^{e}$ , 
Eve Vavagiakis$^{d}$ , 
John Vivalda$^{e}$ , 
Benjamin Westrook$^{b,c}$ , 
Daniel Yohannes$^{e}$ 
\skiplinehalf
\small{
$^{a}$ Physics Division, Lawrence Berkeley National Laboratory, Berkeley, CA 94720, USA
\\$^{b}$ Department of Physics, University of California, Berkeley, CA 94720, USA
\\$^{c}$ Radio Astronomy Laboratory, University of California, Berkeley, CA 94720, USA
\\$^{d}$ Physics Department, Cornell University, Ithaca, NY, 14853, USA
\\$^{e}$ SeeQC, Inc., Elmsford, NY, 10523, USA
}
\skiplinehalf
\small{
$^{*}$ Corresponding author: asuzuki@lbl.gov % e-mail
}
}

%\date{09.30.2015}
%\linenumbers
 \maketitle 

%%%%%%%%%%%%%%%%%%%%%%%%%%%%%% Abstract %%%%%%%%%%%%%%%%%%%%%%%%%%%%%%
\begin{abstract}
We report on the development of commercially fabricated multi-chroic antenna coupled Transition Edge Sensor (TES) bolometer arrays for Cosmic Microwave Background (CMB) polarimetry experiments. The orders of magnitude increase in detector count for next-generation CMB experiments requires a new approach in detector wafer production to increase fabrication throughput. 

We describe collaborative efforts with a commercial superconductor electronics fabrication facility (SeeQC, Inc.) to fabricate antenna coupled TES bolometer detectors. We have successfully fabricated an operational dual-polarization, dichroic sinuous antenna-coupled TES detector array on a 150 mm diameter wafer. The fabricated detector arrays have average yield of 95\% and excellent uniformity across the wafer. Both RF characteristics and TES bolometer properties are suitable for CMB observations. We successfully fabricated different types of TES bolometers optimized for frequency-multiplexing readout, time-domain multiplexing readout, and microwave SQUID multiplexing readout. We also demonstrated high production throughput. We discuss the motivation, design considerations, fabrication processes, test results, and how industrial detector fabrication could be a path to fabricate hundreds of detector wafers for future CMB polarimetry experiments.

\end{abstract}

%%%%%%%%%%%%%%%%%%%%%%%%%%%%%%
%\vspace{-4 mm}
\section{Introduction}
%\vspace{-4 mm}
Over the past two decades, teams from around the world have made increasingly sensitive measurements of the Cosmic Microwave Background (CMB) using telescopes on the ground, balloons and satellites. 
The CMB's uniformity has confirmed the hot big bang model of the universe, and detailed measurements of its small ($10^{-5}$) non-uniformity have led to tight constraints on the composition, geometry, and evolution of the universe. 
The CMB is also weakly polarized through Thomson scattering by free electrons, and this polarization provides access to information that is inaccessible from the temperature anisotropy data \cite{S4ScienceBook}.

To access the wealth of information encoded in the CMB polarization, scientists have deployed experiments designed to characterize the polarization state of the CMB with unprecedented sensitivity. 
Transition Edge Sensor (TES) bolometer detectors for CMB experiments have reached photon-noise limited performance. 
Therefore, the overall sensitivity is increased by scaling up the number of detectors. 
Deployed CMB polarization experiments are grouped in stages:  
Stage-2 experiments deployed in the early 2010's with O(1,000) detectors and have successfully detected B-mode polarization in the CMB that arises from weak gravitational lensing of E-mode polarization; Stage-3 experiments deployed in the late 2010's with O(10,000) detectors. 

Next-generation CMB polarization experiments are in the construction and conceptual study phase. 
The Simons Observatory experiment will deploy in the Atacama desert in Chile in 2020 with $\approx$80,000 detectors distributed between one large aperture (6 meter primary mirror) telescope and three small aperture ($\approx$ 0.5 meter) telescopes. 
At the same time, a Stage-4 ground based CMB experiment, CMB-S4, is being planned to make a definitive measurement of CMB polarization from the ground.
A reference design for this next generation CMB experiment, has been studied by the CMB communities \cite{DSR,CDT, LBNLStudy}, calls for $\approx$500,000 detectors spread across 20 GHz to 300 GHz on three large aperture (6 meter) telescopes and 18 small aperture ($\approx$ 0.5 meter) telescopes deployed at Chile and South Pole sites. 
The \textit{CMB-S4 Science Case, Reference Design, and Project Plan} predict that this next generation CMB experiment configuration will be able to measure $r \leq 0.001$ at 95\% c.l., $\Delta N_{eff} < 0.06$ at 95\% c.l., and produce a mm-wave legacy survey that will advance our understanding of galaxy clusters, gamma-ray bursts, and much more. 

The \textit{CMB-S4 Science Case, Reference Design, and Project Plan} also reported that it will require 432 science grade detector wafers fabricated over three years. 
This requirement is approximately an order of magnitude more detector wafers than that required for all Stage-3 experiments combined. 
As such, microfabrication, assembly, and testing of detector modules for CMB-S4 will be an immense task. 
In order to tackle the production throughput challenge for next-generation CMB experiments, we collaboratedwith commercial microfabrication foundries, SeeQC, Inc.\footnote{SeeQC, Inc. spun off from Hypres, Inc. in 2019} and STAR Cryoelectronics Inc., to utilize their industrial scale production throughput. 

We initiated a technology transfer with SeeQC, Inc. and STAR Cryoelectronics Inc. in late 2015, leading to the successful fabrication of an operational dual-polarization, dichroic sinuous antenna-coupled TES detector array on a 150 mm diameter wafer at both foundaries. 
The performance of these earlier prototypes was reported in Suzuki et al.\cite{SuzukiLTD17}. 
In this report, we discuss subsequent progress made with SeeQC, Inc., highlights of which include: 
\begin{itemize}
\item Transfer of all fabrication steps to SeeQC, Inc., 
\item Demonstration of detector designs that achieve detector performance suitable for ground based CMB experiments
\item Optimization of the silicon nitride film to improve optical efficiency
\item Coupling commercially fabricated detectors to three readout technologies used by CMB experiments
\item High throughput fabrication of a deployable detector array 
\item High yield (95+\%) and uniformity of batch fabricated wafers characterized by automated metrology tools 
\end{itemize}
We discuss the motivation, design considerations, fabrication processes, test results, and how industrial detector fabrication could be a path to fabricate hundreds of detector wafers for future CMB polarimetry experiments.

%The fabricated detector array has yield of over 96\% and excellent uniformity across the wafer. We have also demonstrated stable detector performance over 4 months. Both RF characteristics and TES bolometer properties are suitable for CMB observations. We successfully fabricated different types of TES bolometers optimized for frequency-multiplexing readout, time-domain multiplexing readout, and microwave SQUID multiplexing readout. We discuss the motivation, design considerations, fabrication processes, test results, and how industrial detector fabrication could be a path to fabricate hundreds of detector wafers for future CMB polarimetry experiments.

%%%%%%%%%%%%%%%%%%%%%%%%%%%%%%
%\vspace{-4 mm}
\section{Design and Fabrication}
%\vspace{-4 mm}

\begin{figure}[!h]
\begin{center}
\includegraphics[width=\textwidth,keepaspectratio]{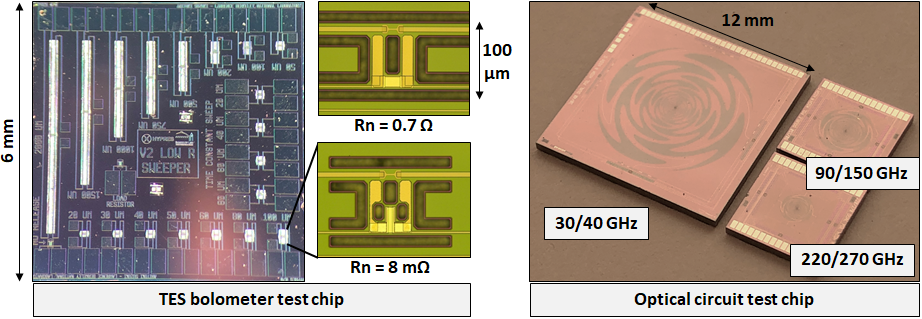}
\end{center}
\caption{[Left] Example of test pixels. Bolometers with different weak link lengths are visible. TES bolometers for different readout electronics are shown as inserts. [Right] Photograph of optical test pixels.
%\vspace{-4 mm}
}
\label{fig:FigA}
\end{figure}

The detector design was based on a sinuous antenna-coupled TES bolometer array fabricated for POLARBEAR/Simons Array and Simons Observatory \cite{WestbrookLTD17, SO2018}. 
The detectors are fabricated on 150 mm silicon wafers that have low stress silicon nitride films pre-deposited by a commercial vendor. 
Next, a Nb/SiN/Nb trilayer is deposited with DC magnetron putter and plasma enhanced chemical vapor deposition processes. 
Films were etched with reactive ion plasma to form RF and bias circuits for the TES bolometers. 
Titanium and palladium films, used to form RF resistors and heat reservoirs for the TES bolometers, respectively, are deposited with e-beam evaporation and patterned using a lift-off process. 
Manganese doped aluminum (AlMn) is sputtered, then patterned with ion-milling to form the TES sensor, which is then passivated with a silicon nitride film to protect it from chemicals such as the photo-resist developer. 
For the bolometer release process, the low stress nitride is first etched to expose the underlying silicon, which is then isotropically etched by xenon-fluoride ($\mathrm{XeF}_2$) gas in order to release the TES bolometer and its low stress nitride support from the substrate. 
Finally, the remaining photo-resist is removed with an oxygen plasma. 
After fabrication, the Wafers are inspected and packaged for further cryogenic characterization. 

In the previous report, two processes, deposition of AlMn film and $\mathrm{XeF}_2$ release, were done at U.C. Berkeley's Marvell NanoLab\cite{SuzukiLTD17}. 
Since then, SeeQC, Inc. procured, installed and tested AlMn targets with three different doping levels (1000 ppm, 2100 ppm and 2500 ppm) at various annealing temperatures, from 160 to 260 degrees Celsius \cite{Dale}. The target critical temperatures were 165 mK and $\approx$500 mK, which were achieved by heating the 2500 ppm target to 215 degrees Celsius and the 2100 ppm target to 200 degrees Celsius, respectively. SeeQC also procured, installed and commissioned a XeF2 machine. These additions to SeeQC's fabrication line enable them to perform all fabrication steps for CMB detector wafers at their foundry, greatly streamlining the fabrication process. The devices fabricated for this report were fabricated end-to-end at SeeQC using the newly acquired AlMn targets and $\mathrm{XeF}_2$ machine.

%%%%%%%%%%%%%%%%%%%%%%%%%%%%%%
%\vspace{-4 mm}
\section{Detector Performance}
%\vspace{-4 mm}
%\vspace{-4 mm}
\begin{figure}[!h]
\begin{center}
\includegraphics[width=\textwidth,keepaspectratio]{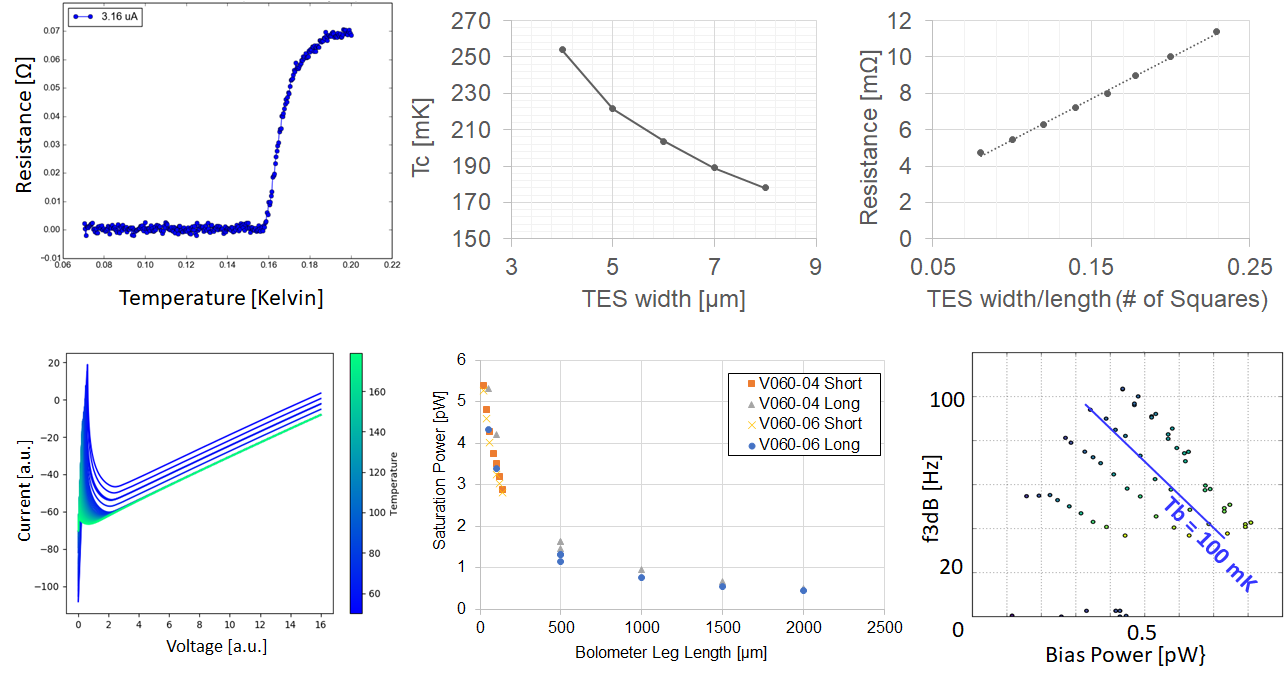}
\end{center}
\caption{[Top] Plots from TES sensor characterization. Superconducting transition of TES (Left), Tc as a function of gaps between Niobium leads that touches TES sensor (Center) and resistance as a function of geometry of TES sensor (Right). [Bottom] Plots from TES bolometer characterization. Current-Voltage curve (Left), saturation power as a function of length of thermal weak links (Center) and time constant as a function of bias power (Right). 
%\vspace{-4 mm}
}
\label{fig:FigG}
\end{figure}

We designed and fabricated test chips, each with several different detector designs to quickly collect data on how detector performance varies as a function of design parameters as shown in Figure \ref{fig:FigA}. 
As an example, the test chip shown in Figure~\ref{fig:FigA} has TES bolometers with various thermal link lengths from 20 microns to 2000 microns. 
The Tests were performed in a dilution refrigerator cryostat using time-domain multiplexing readout. 
Measurements of various TES bolometers properties are shown in Figure~\ref{fig:FigG}. 
Superconducting transition temperature (Tc) of Aluminum-Manganese (2500 ppm manganese) film deposited with DC magnetron sputter and annealed to 215 C exibit Tc of 165 mK that meet CMB-S4 target. 
When we connect AlMn film to Niobium leads, we noticed shifts in Tcs as a function of gap between two niobium leads. This is expected behavior as niobium has high Tc (approximately 9 Kelvin), therefore Tc of AlMn is modified by proximity effect. 
We also explored how resistance of AlMn TES sensor changed as a function of dimensions. Resistance scaled linearly as a functino of number of squares as expected. 

There are three multiplexing readout technologies used by mm (CMB) and sub-mm (IR) wave experiments: frequency multiplexing (fmux), time domain multiplexing (tmux), and microwave-SQUID multiplexing ($\mu$mux). 
The impedance of the TES sensors used are two orders of magnitude different between fmux ($R_n = 0.7\Omega$) and tmux/$\mu$mux ($R_n = 0.008\Omega$). 
To match to the different readouts, We fabricated TES bolometers with different designs as shown in Figure~\ref{fig:FigA}. 
Simply changing lateral dimensions was not enough to vary resistances by two orders of magnitude. 
Therefore, we also changed the thickness of AlMn films by a factor of ten between wafers fabricated for fmux readout and wafers fabricated for tmux/$\mu$mux readouts. 
We successfully fabricated both types of wafers, and detectors were read out by the corresponding fmux, tmux and $\mu$mux readout electronics successfully. 
Tests showed expected bias voltage to readout current relationship as shown in Figure~\ref{fig:FigG} (tmux example shown).  

Varying thermal conductance of thermal weak links exhibit the expected inverse length relationship with saturation powers as shown in Figure~\ref{fig:FigG}.
A plot in the Figure~\ref{fig:FigG} shows saturation powers from different wafers. 
Measurements were repeatable for both different chips from same wafers and different chips from different wafers. 
Our designs covered saturation powers from 0.5 pW to 5.5 pW with a base temperature of 100 mK and a superconducting transition temperature of 165 mK for the TES sensors. This range is a suitable for low (30/40 GHz) and mid (90/150 GHz) frequency detectors for ground based CMB experiments. 

We also measured time constant of a bolometer at various bath temperature by looking at time response of TES bolometer to small change in electrical bias power. 
In the Figure~\ref{fig:FigG}, we highlighted measurement done at 100 mK. 
Our bolometer had response speed of 40 Hz to 100 Hz. Detector sped up as TES sensor is dropped into superconducting transition as expected. 
We deposited palladium thickness of 1 micron as a thermal anchor during fabrication of the TES bolometer. 
Thickness of palladium can be adjusted to speed up/down TES sensors. 

These tests demonstrate that SeeQC, Inc. is capable of fabricating detectors for all readout types used by CMB experiments. 

%\vspace{-4 mm}
\begin{figure}[!h]
\begin{center}
\includegraphics[width=\textwidth,keepaspectratio]{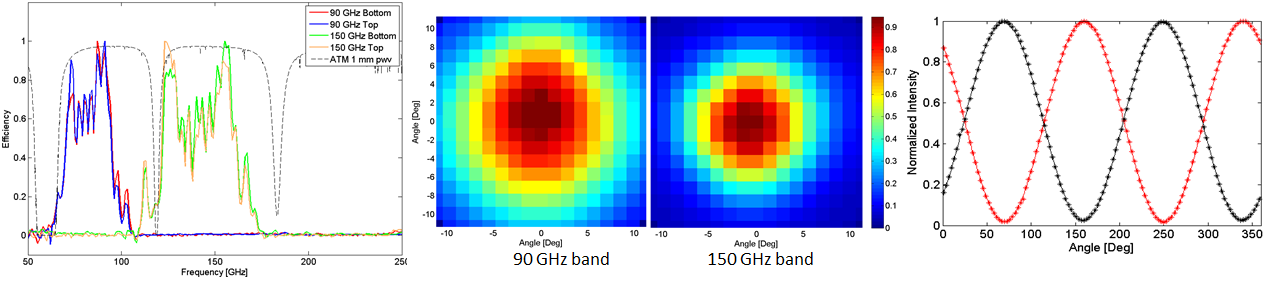}
\end{center}
\caption{Plots from optical tests. Spectra (Left), beam (Center) and polarization (Right). 
%\vspace{-4 mm}
}
\label{fig:FigH}
\end{figure}

We have also fabricated optical test pixels as shown in Figure~\ref{fig:FigA} to optimize the RF performance of the detectors. 
While we have fabricated low (30/40 GHz), mid (90/150 GHz), and high (220/280 GHz) dual frequency band detectors, we have so far focused on mid-frequency detectors for characterization. 

Characterization of detectors at cryogenic temperature was conducted in an eight-inch Infrared Labs wet dewar with a Helium-3 adsorption fridge that achieves 250-mK base temperatures. A brief description of the optical test setup is given here. A detailed description can be found elsewhere\cite{SuzukiThesis}.
We mounted the detector chip on an anti-reflection coated synthesized elliptical alumina lens. 
The dewar is equipped with Quantum Design DC SQUID readout electronics. 
Voltage biases to TES bolometers were provided by a battery at room temperature connected to a 20 milli-Ohm shunt resistor mounted on the 4 Kelvin stage.
The dewar has a zotefoam optical window with metal mesh low pass filters provided by Cardiff university. 
A Michelson Fourier transform spectrometer with a 254 micron thick mylar beam splitter was used to characterize the spectral response.
The beam performance was characterized by scanning a 12.7 mm diameter temperature modulated source in front of the dewar window on a 2-D linear actuated stage.
Polarization response was characterized by rotating a wire grid in front of the temperature modulated source located at the peak of the antenna angular response. 

Test results are shown in Figure~\ref{fig:FigH}.
Spectral measurements show that the signal from the broadband sinuous antenna is split into 90 GHz and 150 GHz frequency bands.
In addition, spectra measurements matched well between the two orthogonal polarization channels. 
However, band passes were shifted down in frequency relative to the design. 
This shift is due to the fact that during design, we used an assumed value for the at the time unknown dielectric constant of the silicon nitride film. 
We have since calculated the dielectric constant information for SeeQC's silicon nitride from the band pass measurements and updated our design accordingly.
Detectors with the updated design are already fabricated and ready for testing. 
Beam measurements show round beams for both the 90 GHz and 150 GHz bands: 
ellipticity, defined as the difference in beam widths, defined as an angular width at 1/e point, divided by their sum, were $3\pm1$\% and $2\pm1$\% for 90 GHz and 150 GHz, respectively. 
Finally, polarization leakage was 2\% and 2\% for 90 GHz and 150 GHz, respectively.

We developed a silicon-rich silicon nitride film to achieve high RF transmission efficiency between the antenna and detector. 
To fabricate detectors, we used silicon nitride with a dielectric constant as high as 8. 
We measured end-to-end (outside of cryostat to TES bolometer) optical efficiency of $\approx$70\% for 90 GHz band and $\approx$50\% for 150 GHz. 

%%%%%%%%%%%%%%%%%%%%%%%%%%%%%%
%\vspace{-4 mm}
\section{Detector Array Fabrication}
%\vspace{-4 mm}
\begin{figure}[!h]
\begin{center}
\includegraphics[width=\textwidth,keepaspectratio]{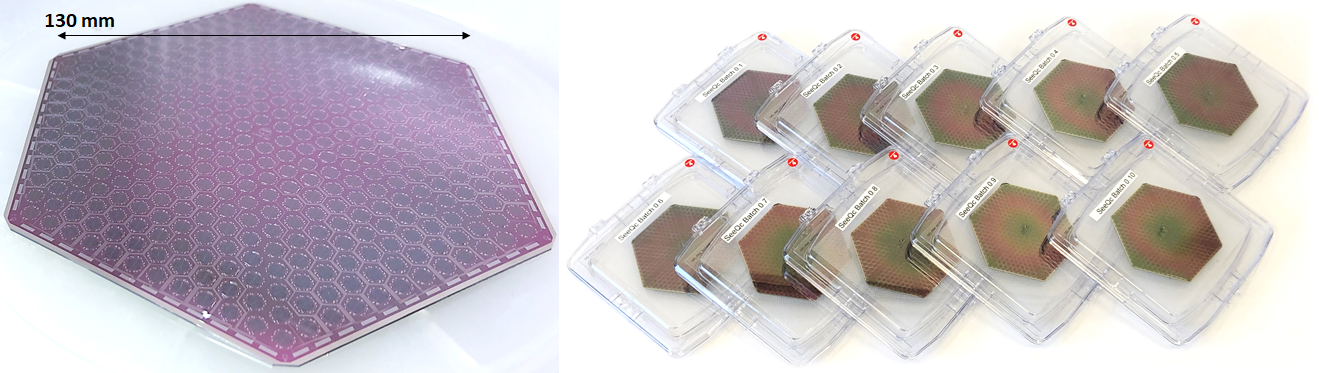}
\end{center}
\caption{Photograph of detector arrays.
%\vspace{-4 mm}
}
\label{fig:FigD}
\end{figure}

Encouraged by the test pixel results, we decided to fabricate a detector array. 
The detector array design was based on the Simons Observatory experiment's mid-frequency detector. 
The fabrication steps for the detector array is identical to that of the test pixels. 
We successfully fabricated a detector array on the first try as shown in Figure~\ref{fig:FigD}. 

We then fabricated ten wafers in a batch to demonstrate throughput at SeeQC, Inc. 
We calculated that fabrication of 10 wafers in 15 working days is the fabrication pace required to meet the demand of CMB-S4. 
To maximize production throughput, we grouped the ten wafers into five groups of two wafers, then staggered fabrication of the five groups by separating them by a day each. 
All ten wafers were delivered successfully in 15 working days, including time for making monitoring samples, metrology, and documentation.
Thus, we achieved our production throughput. 

During the batch fabrication, we identified a bottleneck in the process: 
the e-beam evaporation step took three times longer than the average of the other steps because the machine could only process one wafer at a time. 
In the future, we will move this step to a high-throughput e-beam evaporator at SeeQC, Inc. that can process nine wafers at a time in order to remove this bottleneck and fabricate even faster. 

%\vspace{-4 mm}
\begin{figure}[!h]
\begin{center}
\includegraphics[width=\textwidth,keepaspectratio]{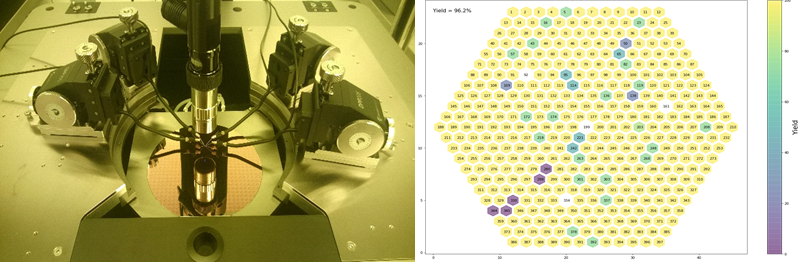}
\end{center}
\caption{Photograph of automated programmable probe station at SeeQC Inc. Example of yield color map is shown. Yellow pixels have all sensors in a pixel working. Purple pixels show all sensors are not working. We noticed that same pixels were highlighted as purple across all wafers. This allowed us to track down yield hit to probing programming error. 
%\vspace{-4 mm}
}
\label{fig:FigE}
\end{figure}

Wafers were characterized at room temperature, using a programmable probe station to automatically probe through $\approx$2000 TES bolometers on a detector wafer as shown in Figure~\ref{fig:FigD}.
The average DC probe yield from ten wafers was 95\%. 
We identified that 1\% of the yield loss was due to design error, 1\% was due to a probe programming error, and 3\% was due to photo-resist imperfection. These losses will be addressed in future batches to improve the yield further. 

During the batch fabrication, we also collected statistics on film properties such as thickness, resistivity, and stress. 
Silicon nitride film thicknesses were within 1\% of the target thickness with a thickness variation of 1\% across each wafer. 
Niobium film thicknesses were within 5\% of the target thickness with a thickness variation of 1.8\% across each wafer. 
AlMn film thicknesses were within 3\% of the target thickness with a sheet resistance variation of less than 5\% across each wafer. 
These values are all well within the detector specifications for next generation CMB experiments. 

%%%%%%%%%%%%%%%%%%%%%%%%%%%%%%
%\vspace{-6 mm}
\section{Conclusion and Future Developments}
%\vspace{-3 mm}
We successfully designed, fabricated, and tested an antenna-coupled TES bolometer array with SeeQC, Inc. 
Performance, yield, and throughput results are encouraging for a commercial foundry to tackle the challenges of detector fabrication for next-generation CMB experiments. 
Cryogenic characterization of a fabricated detector arrays is an important next step to raise the technology readiness level of the idea. 

%%%%%%%%%%%%%%%%%%%%%%%%%%%%%%
%\vspace{-4 mm}
\begin{acknowledgements}
This work was supported by: Early Career Research Program, Office of Science, of the U.S. Department of Energy under Contract No. DE-AC02-05CH11231. Small Business Innovation Research (SBIR), Office of Science, of the U.S. Department of Energy under Award No. DE-SC0017818 and DE-SC0018711. Simons Observatory. 
\end{acknowledgements}

%%%%%%%%%%%%%%%%%%%%%%%%%%%%%%
\clearpage
%\bibliographystyle{unsrt}
%\bibliography{LTD18ReferenceShort} 

\begin{thebibliography}{1}

\bibitem{S4ScienceBook}
Kevork~N. Abazajian et~al.
\newblock {CMB-S4 Science Book, First Edition}.
\newblock {\em arXiv: 1907.04473} [astro-ph.IM], Oct 2016.

\bibitem{DSR}
Kevork Abazajian et~al.
\newblock {CMB-S4 Science Case, Reference Design, and Project Plan}.
\newblock {\em arXiv: 1907.04473 [astro-ph.IM]}, Jul 2019.

\bibitem{CDT}
Concept Definition~Task Force.
\newblock {Cosmic Microwave Background Stage 4 Concept Dedinition Task Force}.
\newblock \url{https://cmb-s4.org/CMB-S4workshops/images/CMBS4_CDT_final.pdf},
  2017.
\newblock [Online; accessed 4-Nov-2017].

\bibitem{LBNLStudy}
Darcy Barron et~al.
\newblock Optimization study for the experimental configuration of {CMB}-s4.
\newblock {\em Journal of Cosmology and Astroparticle Physics},
  2018(02):009--009, Feb 2018.

\bibitem{SuzukiLTD17}
Aritoki Suzuki et~al.
\newblock Commercialization of micro-fabrication of antenna-coupled transition
  edge sensor bolometer detectors for studies of the cosmic microwave
  background.
\newblock {\em JLTP}, 193(5):744--751, Dec 2018.

\bibitem{WestbrookLTD17}
Benjamin. Westbrook et~al.
\newblock The polarbear-2 and simons array focal plane fabrication status.
\newblock {\em JLTP}, 193(5):758--770, Dec 2018.

\bibitem{SO2018}
Nicholas Galitzki et~al.
\newblock The simons observatory: instrument overview.
\newblock {\em SPIE Proceedings, Millimeter, Submillimeter, and Far-Infrared
  Detectors and Instrumentation for Astronomy IX}, 10708, Jul 2018.

\bibitem{Dale}
Dale Li et~al.
\newblock Almn transition edge sensors for advanced actpol.
\newblock {\em Journal of Low Temperature Physics}, 184(1):66--73, Jul 2016.

\bibitem{SuzukiThesis}
Aritoki Suzuki.
\newblock {\em {Multichroic Bolometric Detector Architecture for Cosmic
  Microwave Background Polarimetry Experiments}}.
\newblock PhD thesis, University of California, Berkeley, Dec 2013.

\end{thebibliography}

\end{document}